\documentclass[aps,pra]{revtex4}

\usepackage{graphicx}
\begin{document}

\title{Uncertainty characteristics of generalized 
quantum measurements}

\author{Holger F. Hofmann}
\email{h.hofmann@osa.org}
\affiliation{PRESTO, 
Japan Science and Technology Corporation 
(JST),\\
Research Institute for Electronic Science, Hokkaido 
University\\
Kita-12 Nishi-6, Kita-ku, Sapporo 060-0812, Japan}

\date{\today}

\begin{abstract}
The effects of any quantum measurement can be described 
by a collection of measurement operators $\{\hat{M}_m\}$ 
acting on the quantum state of the measured system. 
However, the Hilbert space formalism tends to obscure the 
relationship between the measurement results and the 
physical properties of the measured system.
In this paper, a characterization of measurement operators
in terms of measurement resolution and disturbance is
developed. It is then possible to formulate uncertainty 
relations for the measurement process that are valid for
arbitrary input states. 
The motivation of these concepts is explained 
from a quantum communication viewpoint. It is shown that 
the intuitive interpretation of uncertainty as a relation
between measurement resolution and disturbance provides a
valid description of measurement back action. Possible 
applications to quantum cryptography, quantum cloning, and 
teleportation are discussed.
\end{abstract}


\maketitle

\section{Introduction}
One of the most intriguing problems of quantum mechanics is
the interpretation of the measurement process \cite{Meas}.
The reason for this central role of the measurement process
is the absence of fundamental "elements of reality" that
would simultaneously characterize both the dynamics and
the measurement results \cite{EPR,Bel64,Mer90}.
It is therefore not possible to
trace the measurement interaction back to microscopic 
trajectories. Instead, only a summary of the total statistical
effects of a measurement is available.
Originally, this property of quantum mechanics was explained by 
Heisenberg in terms of an uncontrollable disturbance in one
variable caused by the measurement of another variable 
\cite{Hei58}.
However, this explanation was still based on a classical model
of the measurement interaction. Consequently, the general 
validity of Heisenberg's original argument has been questioned
by a number of researchers \cite{Nie,Scu91}. 
In particular, 
there appear to be some unresolved issues concerning the
derivation of uncertainties using correlations between the 
system and the measurement device \cite{Art88,Oza02a,Oza02b}.

On the other hand, the investigation of various methods to 
prepare and control quantum states, especially in the 
field of quantum optics, has motivated the development 
of a generalized measurement theory based on the Hilbert 
space representation of quantum states. This formalism 
allows an expression of all relevant statistical properties 
of a quantum measurement in an extremely compact form 
\cite{NieMT}. Unfortunately, this compact form tends to 
obscure the relationship between physical properties of 
the system and the measurement process. 
In particular, the relationship of this generalized formulation 
of measurement with Heisenberg's original discussion of the
uncertainty principle as a relation between measurement
resolution and the disturbance of a conjugate variable may not
be entirely clear \cite{Nie}. 

In this paper, 
the measurement effects of a generalized measurement described by 
a set of operators $\{\hat{M}_m\}$ are characterized in terms of
the physical properties of the measured system. This 
characterization is based on the reliability of quantitative 
estimates for various physical properties of the system before 
the measurement. Using these definitions, the uncertainty relations 
for measurement resolution and disturbance can be derived, 
thereby establishing the validity of the uncertainty principle 
for generalized quantum measurements. It is then possible to 
translate Heisenberg's original argument into a form closer to 
present problems in quantum information theory. In particular, 
it is shown that the concept of disturbance can be understood 
in terms of a loss of information about the input state
caused by the measurement back action. This interpretation 
can then be applied to problems such as quantum cryptography, 
quantum cloning, and quantum teleportation.

\section{Quantitative estimates and measurement resolution}
\label{sec:quant}

While classical physics allows a direct identification of
measurement results with objective properties of the 
system, the existence of which is thought to be independent
of the measurement process, quantum mechanics is formulated in 
an abstract probabilistic space from which the measurement
statistics must be derived indirectly. Therefore, a special
theory is necessary to identify and define the connection 
between a measurement outcome $m$ and the quantum state of 
the system. In general, this can be achieved by using a set
of measurement operators $\{\hat{M}_m\}$. These operators
describe both the measurement probabilities $p(m)$ for
each outcome $m$ and the change of the quantum state
caused by the measurement back action. For an input density
operator $\hat{\rho}_{\,\mbox{in}}$, 
\begin{eqnarray}
p(m) &=& \mbox{tr}\{\hat{\rho}_{\,\mbox{in}} 
\hat{M}_m^\dagger \hat{M}_m \}
\nonumber 
\\
\hat{\rho}_{\,\mbox{out}} &=& \frac{1}{p(m)} 
\hat{M}_m\hat{\rho}_{\,\mbox{in}}\hat{M}_m^\dagger.
\end{eqnarray}
Note that the properties of $\hat{M}_m$ are only restricted
by the fact that the sum of all probabilities must be one
for any input state, that is
\begin{equation}
\sum_m \hat{M}_m^\dagger \hat{M}_m = \hat{1}.
\end{equation}

In this formulation, the information obtained about the measured
system is represented by the dependence of the measurement probability
$p(m)$ on the input state $\hat{\rho}_{\mbox{\, in}}$. 
In order to characterize the measurement information obtained about 
an observable $\hat{A}$, it is necessary to examine how the 
probability $p(m)$ varies for different eigenstates of $\hat{A}$. 
Suppose that the input state is an unknown eigenstate of the
observable $\hat{A}$. It is then possible to estimate the
eigenvalue of $\hat{A}$ based on the measurement result $m$.
Assuming that each eigenstate of $\hat{A}$ is equally 
likely to be the input, the probability $p(A|m)$ that $m$ 
was obtained as a result of $A$ is given by
\begin{eqnarray}
p(A|m) &=& \frac{
\langle A \mid \hat{M}_m^\dagger\hat{M}_m \mid A \rangle}
{\sum_A^\prime 
\langle A^\prime \mid \hat{M}_m^\dagger\hat{M}_m\mid A^\prime \rangle}
\nonumber \\
&=&
\frac{
\langle A \mid \hat{M}_m^\dagger\hat{M}_m \mid A \rangle}
{\mbox{tr}\{\hat{M}_m^\dagger\hat{M}_m\}}.
\end{eqnarray}
In order to provide a single quantitative estimate of the
input eigenvalue $A$, it is necessary to assign a 
measurement value $A_m$ to each possible outcome $m$.
The reliability of this estimate can be characterized by 
the average quadratic error obtained from the probabilities
$p(A|m)$,
\begin{eqnarray}
\label{eq:error}
\delta A_m^2 &=& \sum_A (A_m-A)^2 p(A|m)
\nonumber \\
&=& \frac{\mbox{tr}\{(A_m-\hat{A})^2 \hat{M}_m^\dagger\hat{M}_m \}}
{\mbox{tr}\{\hat{M}_m^\dagger\hat{M}_m\}}.
\end{eqnarray}
The best possible estimate is then obtained by minimizing this
quadratic error. The result of this optimization is the
average value of $A$ in the input state distribution $p(A|m)$,
\begin{eqnarray}
\label{estav}
A_m &=& \sum_A A \; p(A|m)
\nonumber \\
&=& \frac{\mbox{tr}\{\hat{A}\; \hat{M}_m^\dagger\hat{M}_m \}}
{\mbox{tr}\{\hat{M}_m^\dagger\hat{M}_m\}}.
\end{eqnarray}
The measurement outcome $m$ can then be identified 
with a quantitative measurement of the observable 
$\hat{A}$, where the measurement result is given by 
$A_m$ and the measurement resolution is given by 
$\delta A_m^2$. 

This procedure can be applied to any observable. 
The measurement result $m$ therefore 
provides information about all physical properties
of the measured system. 
The information obtained from the
measurement result $m$ can be summarized by
the normalized statistical measurement operator
$\hat{R}_m$ given by
\begin{equation}
\label{eq:Rm}
\hat{R}_m = \frac{\hat{M}^\dagger_m \hat{M}_m}
{\mbox{tr}\{\hat{M}_m^\dagger\hat{M}_m\}}.
\end{equation}
This operator is essentially a time-reversed version of the
density matrix. Instead of predicting future measurement
results, it is used to ``retrodict'' properties of the input
\cite{Bar00,Bar01}. In particular, the quantitative estimates
$A_m$ and the measurement errors $\delta A_m^2$ are now
defined as expectation values of the operator and fluctuations
of the statistical operator $\hat{R}_m$,
\begin{eqnarray}
\label{eq:estimate}
A_m &=& \mbox{tr}\{\hat{A} \; \hat{R}_m\}
\nonumber \\
\delta A_m^2 &=& \mbox{tr}\{\hat{A}^2 \; \hat{R}_m\} - A_m^2.
\end{eqnarray}
The analogy between the statistical operator $\hat{R}_m$
and the density operator indicates that the same uncertainty
relations that apply to quantum state preparation also 
apply to the simultaneous measurement of non-commuting 
properties (see the appendix for a general derivation
of uncertainty relations). Specifically, the uncertainty limit of 
the measurement errors $\delta A_m^2$ and $\delta B_m^2$ of 
two non-commuting observables $\hat{A}$ and $\hat{B}$ 
is given by
\begin{equation}
\label{eq:uncertainty1}
\delta A_m^2 \delta B_m^2 \geq 
\frac{1}{4} |\mbox{tr}\{ \hat{R}_m \; [\hat{A},\hat{B}]\}|^2. 
\end{equation}
This uncertainty relation applies to cases where the
same measurement procedure is used to estimate both input
eigenvalues of $\hat{A}$ and input eigenvalues of $\hat{B}$
when no additional information on the input state is 
available.
An example of such a situation can be given in terms of a
quantum cryptography protocol, where a message can be either 
encoded in the eigenvalues of $\hat{A}$ or in the eigenvalues of 
$\hat{B}$. The uncertainty relation (\ref{eq:uncertainty1}) 
then defines a quantitative limitation on eavesdropping 
attempts. An important feature of this measurement uncertainty
is that it does not depend on the input state at all. 
Instead, the uncertainty limit is defined
by an expectation value of the statistical matrix $\hat{R}_m$
that characterizes the information obtained in the
measurement. In general, this expectation value can itself 
be interpreted as an estimate of a physical
property of the measured system. The characterization of
measurement uncertainty is thus achieved entirely in 
terms of information obtained in the measurement,
avoiding any ambiguities of assumptions about the physical 
reality represented by the input state. Nevertheless the 
uncertainty relation (\ref{eq:uncertainty1}) also has 
implications for the interpretation of the measurement 
back action, as will be 
explained in the next section.

\section{Back action and disturbance}
The measurement operator $\hat{M}_m$ not only describes 
the measurement information obtained from the measurement
result $m$, but also the measurement back action effects
that change the input state into the output state.
In order to characterize this change in terms of physical
properties, it is necessary to find a useful definition 
of the disturbance of an observable $\hat{B}$ caused by
the measurement $\hat{M}_m$. In the following, the 
definition of disturbance will be based on the measurement
error obtained when the input eigenvalue of $\hat{B}$ is
estimated by the outcome of a precise projective measurement
of $\hat{B}$ performed on the output of the measurement
$\hat{M}_m$. The disturbance then corresponds to a loss
of information about the property $\hat{B}$ suffered as
a consequence of the measurement back action.

The measurement sequence defined by an initial measurement 
result $m$ followed by a final measurement result of 
$B_f$ is characterized by a statistical matrix $\hat{R}_{mf}$
given by
\begin{eqnarray}
\label{eq:rmf}
\hat{R}_{mf} &=& \mid r_{mf} \rangle \langle r_{mf} \mid
\nonumber \\ &&
\mbox{with} \hspace{0.3cm} 
\mid r_{mf} \rangle =  
\frac{1}{\sqrt{
\langle B_f \mid \hat{M}_m\hat{M}_m^\dagger \mid B_f \rangle}}
\hat{M}_m^\dagger \mid B_f \rangle
.
\end{eqnarray}
This statistical matrix determines the optimal estimates
for the input eigenvalues of both $\hat{A}$ and $\hat{B}$
obtained from the measurement results $m$ and $B_f$,
\begin{eqnarray}
A_{mf} &=& \langle r_{mf} \mid  \; \hat{A} \; \mid r_{mf} \rangle
\nonumber \\
B_{mf} &=& \langle r_{mf} \mid  \; \hat{B} \; \mid r_{mf} \rangle.
\end{eqnarray}
The measurement errors are given by
\begin{eqnarray}
\delta A_{mf}^2 &=&
\langle r_{mf} \mid  \; \hat{A}^2 \; \mid r_{mf} \rangle
- A_{mf}^2
\nonumber \\
\delta B_{mf}^2 &=&
\langle r_{mf} \mid  \; \hat{B}^2 \; \mid r_{mf} \rangle
- B_{mf}^2.
\end{eqnarray}
The proper uncertainty relation for the estimates of 
of $\hat{A}$ and $\hat{B}$ obtained from an initial 
measurement $\hat{M}_m$ followed by a precise measurement 
of $\hat{B}$ therefore reads
\begin{equation}
\delta A_{mf}^2 \delta B_{mf}^2 \geq
\frac{1}{4}
\left|
\langle r_{mf} \mid  \; [\hat{A},\hat{B}] \; 
\mid r_{mf} \rangle
\right|^2.
\end{equation}
Since this uncertainty relation applies to the best
estimates of $\hat{A}$ and $\hat{B}$, it is obvious that 
it is also fulfilled for estimates of $\hat{A}$ based only
on $m$ and estimates of $\hat{B}$ given directly by the 
final measurement value $B_f$. 
In particular, the actual disturbance $\Delta B_{mf}^2$
can be written as
\begin{eqnarray}
\Delta B_{mf}^2 &=& \langle r_{mf} \mid (B_f-\hat{B})^2 
\mid r_{mf} \rangle
\nonumber \\
&=& \delta B_{mf}^2 + (B_f - B_{mf})^2,
\end{eqnarray}
where the difference between the measurement result $B_f$
and the optimal estimate $B_{mf}=\langle r_{mf} \mid \hat{B}
\mid r_{mf} \rangle$ corresponds to a systematic error
caused by the measurement back action. The disturbance can 
thus be separated into a random part $\delta B_{mf}^2$
limiting the possibility to estimate the initial value
of $\hat{B}$, and a well defined shift of $\hat{B}$ 
given by the difference between the output value $B_f$ and
the best possible estimate $B_{mf}$ of the unknown input 
value based on all the available measurement information.

With the definitions given above, it is possible to 
formulate Heisenberg's uncertainty principle with respect
to the measurement resolution $\delta A_{mf}$ and the
disturbance $\Delta B_{mf}^2$ caused by the measurement 
back action. In this form, it states that the error 
$\delta A_{mf}^2$ of an estimate of $\hat{A}$ obtained from
the measurement result $m$ is related to the disturbance 
$\Delta B_{mf}^2$ of a non-commuting property $\hat{B}$
by
\begin{equation}
\label{eq:mfrelation}
\delta A_{mf}^2 \Delta B_{mf}^2 \geq
\frac{1}{4}
\left|
\langle r_{mf} \mid  \; [\hat{A},\hat{B}] \; 
\mid r_{mf} \rangle
\right|^2.
\end{equation} 
This relation takes into account the final measurement
result $B_f$ and may include correlations between the
error of the estimate $A_m$ and the final result
$B_f$. Therefore, a complete characterization of 
$\hat{M}_m$ requires the determination of measurement
resolutions $\delta A_{mf}$ and disturbances 
$\hat{B}_{mf}$ for each final result $B_f$.

In order to obtain a single expression for the disturbance
of $\hat{B}$ caused by the measurement $\hat{M}_m$, it
is necessary to average over all final measurement results 
$B_f$. For this purpose, 
the statistical matrix $\hat{R}_m$ for the measurement
of $m$ given by equation (\ref{eq:Rm}) can be expressed in terms
of the eigenstate $\mid r_{mf} \rangle$ of the statistical
matrix $\hat{R}_{mf}$ for the joint measurement given by 
equation (\ref{eq:rmf}), 
\begin{eqnarray}
\label{eq:decomp}
\hat{R}_m &=& \sum_{B_f} 
w_m (B_f)
\mid r_{mf} \rangle \langle r_{mf} \mid
\nonumber \\
&& \mbox{with} \hspace{0.3cm}
w_m (B_f) =
\frac{\langle B_f \mid \hat{M}_m 
\hat{M}_m^\dagger \mid B_f \rangle}
{\mbox{tr} \{\hat{M}_m^\dagger \hat{M}_m \}}
. 
\end{eqnarray}
This decomposition shows that the statistical
matrix $\hat{R}_m$ can be interpreted as an
average over the statistical matrices $\mid r_{mf}\rangle
\langle r_{mf} \mid$ of each possible final outcome $B_f$
with the appropriate statistical weights $w_m(B_f)$.
If no other information on the input state is available,
$w_m(B_f)$ is the conditional probability of obtaining
the final result $B_f$ following an initial measurement 
result of $m$. 
Since the errors of the estimate $A_m$ obtained from the
measurement result $m$ are given by 
an expectation value of the statistical matrix, it 
follows from equations (\ref{eq:decomp}) and 
(\ref{eq:error}) that 
\begin{eqnarray}
\delta A_m^2 &=& \mbox{tr} \{(A_m-\hat{A})^2 \; \hat{R}_m \}
\nonumber \\
&=& \sum_{B_f} w_m(B_f) 
  \langle r_{mf} \mid(A_m-\hat{A})^2 \mid r_{mf} \rangle
\nonumber \\
&=& \sum_{B_f} w_m(B_f) \delta A_{mf}^2.
\end{eqnarray}
The measurement error $\delta A_m^2$ obtained for $\hat{M}_m$
is therefore equal to the statistical average over the 
measurement errors $\delta A_{mf}^2$ obtained for the measurement
sequences $\langle B_f \mid \hat{M}_m$.
Likewise, the averaged disturbance of $\hat{B}$ associated
with the measurement result $m$ can be obtained by 
\begin{eqnarray}
\label{eq:distall}
\Delta B_m^2 &=& \sum_{B_f} w_m(B_f) \Delta B_{mf}^2
\nonumber \\
&=& \sum_{B_i,B_f} w_m(B_f) |\langle r_{mf}\mid B_i \rangle|^2
(B_f-B_i)^2
\nonumber \\
&=& \sum_{B_i,B_f} 
\frac{|\langle B_f\mid \hat{M}_m \mid B_i \rangle|^2}
{\mbox{tr}\{\hat{M}_m^\dagger \hat{M}_m\}} (B_f-B_i)^2.
\end{eqnarray}
This definition of measurement disturbance corresponds to 
an average of the squared difference between the final
value $B_f$ and the initial value $B_i$ over all possible 
input and output values of $\hat{B}$. This average can 
be obtained experimentally and corresponds well with the
intuitive idea of disturbance as a random change of 
$\hat{B}$. 

Since $\delta A_m^2$ and $\Delta B_m^2$ can both be expressed
as averages over $\delta A_{mf}^2$ and $\Delta B_{mf}^2$,
it is now possible to derive an uncertainty relation
for $\delta A_m^2$ and $\Delta B_m^2$ from the relations
(\ref{eq:mfrelation}) for each 
$\delta A_{mf}^2$ and $\Delta B_{mf}^2$. As shown in the 
appendix, the uncertainty of a statistical mixture can
be derived directly from the individual uncertainties by
averaging the corresponding uncertainties as well,
\begin{equation}
\label{eq:mix}
\left(\sum_{B_f} w_m(B_f) \delta A_{mf}^2\right) 
\left(\sum_{B_f} w_m(B_f) \Delta B_{mf}^2\right)
\geq
\frac{1}{4}
\left(\sum_{B_f} w_m(B_f) \left|
\langle r_{mf} \mid  \; [\hat{A},\hat{B}] \; 
\mid r_{mf} \rangle
\right| \right)^2.
\end{equation}
The uncertainty limit on the right side of the equation
can be simplified by noting that
\begin{equation}
\left(\sum_{B_f} w_m(B_f) |\langle r_{mf} \mid
\; [\hat{A}, \hat{B}] \; \mid r_{mf} \rangle| \right)^2
\geq 
\left|\mbox{tr}\{\hat{R}_m\;[\hat{A},\hat{B}]\}\right|^2.
\end{equation}
With this simplification, it is now possible to formulate
the uncertainty given by (\ref{eq:mix}) without any
explicit sums over the final results $B_f$.
For any measurement described by a measurement operator
$\hat{M}_m$, the measurement error $\delta A_m^2$ of the
best estimate of $\hat{A}$ obtained from $m$ and the 
disturbance $\Delta B_m^2$ in the property $\hat{B}$ 
caused by the measurement back action of $\hat{M}_m$
obey the uncertainty relation
\begin{equation}
\label{eq:uncertainty2}
\delta A_m^2 \Delta B_m^2 \geq 
\frac{1}{4} |\mbox{tr}\{ \hat{R}_m \; [\hat{A},\hat{B}]\}|^2. 
\end{equation}
This limit shows that the uncertainty principle does
indeed apply to the relation between measurement resolution
and disturbance, contrary to the statement found in the 
otherwise excellent book by Nielsen and Chuang \cite{Nie}.
Moreover, it suggests that reports on possible violations 
of measurement uncertainty \cite{Scu91,Oza02a,Oza02b,Dur98} 
are based on definitions of measurement resolution and 
disturbance that are not consistent with the ones given 
here. The definition of uncertainties in terms of the
information obtained about unknown input states given
in equations (\ref{eq:error}) and (\ref{eq:distall}) may 
therefore be closer to the original intention of Heisenberg's
argument than the alternatives.

A significant feature of the uncertainty relation
(\ref{eq:uncertainty2}) is that it characterizes the actual
changes in a physical property caused by the measurement
given by the disturbance $\Delta B_m^2$.
This disturbance is given in terms of information that 
may be available before and after the measurement, 
but it does not directly refer to the information obtained
in the measurement process itself. By relating this 
disturbance in $\hat{B}$ to the measurement resolution 
in $\hat{A}$, the uncertainty (\ref{eq:uncertainty2}) 
establishes an inseparable connection between physics 
and information that may be one of the most 
characteristic features of quantum mechanics.
Consequently, a complete characterization of quantum 
measurements must always include both the information 
aspect given by the measurement resolution and the 
dynamical aspect given by the disturbance.
 
Two simple examples of photon number measurements 
may help to illustrate the different aspects
of measurements expressed by resolution and disturbance.
Conventional photon detection usually requires the 
absorption of all photons. The detection of a single 
photon can therefore be represented by the operator
$\hat{M}_{n=1}=\mid n=0 \rangle\langle n=1 \mid$. This 
operator has a perfect measurement resolution of 
$\delta n^2=0$, but its disturbance is given by $\Delta n^2=1$. 
On the other hand, a quantum nondemolition measurement of
photon number is represented by a measurement operator
$\hat{M}_m= \sum_n M_{m}(n) \mid n \rangle\langle n \mid$.
This operator commutes with $\hat{n}$ and therefore has
a disturbance of $\Delta n^2=0$. However, the coefficients 
$M_{m}(n)$ are usually given by a slowly varying function 
of $n$ and the corresponding measurement resolution is very low 
($\delta n \gg 1$). These examples show that the measurement
resolution and the disturbance of a single property are not
usually connected in any way. Interestingly, the uncertainty
(\ref{eq:uncertainty2}) does establish such a connection for
non-commuting properties.

\section{Application to problems in quantum communication}

The application of quantitative concepts to quantum 
communication may appear to be a bit unusual.
Theoretically, it does not make a difference whether
the eigenvalue difference of two orthogonal states 
used in a quantum code is large or small. 
However, the quantitative aspect may be reintroduced by
the specific physical implementation. In multi level 
systems, a reasonable choice of operator properties will 
then represent the fact that weak interactions with the 
environment are more likely to cause transitions between
eigenstates if the eigenvalue difference is small. 
In the presence of noise, it is then optimal to encode 
information in such a way that the more likely errors 
causing small changes in the eigenvalues of $\hat{A}$ 
or $\hat{B}$ are less serious than the comparatively 
unlikely errors involving large changes. Such codes 
will have a quantitative character similar to that of
analog signals. 
In fact, this kind of situation is well known in the
case of continuous variable quantum optics, where
the concept of uncertainty can be applied directly to
implementations of quantum cryptography \cite{Ral00,Hof01}.

A quantum cryptography protocol for the general case
of non-commuting variables $\hat{A}$ and $\hat{B}$ may 
be implemented as follows. Alice will randomly
choose either an eigenstate of $\hat{A}$ or an eigenstate
of $\hat{B}$ to send her information. Likewise, Bob chooses
randomly whether to measure $\hat{A}$ or $\hat{B}$.
By later exchanging data on their choices of $\hat{A}$ or
$\hat{B}$, they can then select the valid communication
attempts. An eavesdropper can now try to optimize the 
simultaneous extraction of information about the eigenvalues 
of $\hat{A}$ and of $\hat{B}$ by choosing various measurement 
strategies $\{\hat{M}_m\}$ with the appropriate resolutions 
$\delta A_m^2$ and $\delta B_m^2$. However, this eavesdropping 
attempt will cause additional noise in the communication 
between Alice and Bob. This noise is given by the disturbances 
$\Delta A_m^2$ and $\Delta B_m^2$ and may lead to the detection 
of the eavesdropper by Alice and Bob. In fact, Alice and Bob
can determine the average disturbances by exchanging information
about the initial eigenvalues sent by Alice and 
the final eigenvalues received by Bob. From randomly selected  
subsets of the valid communication attempts, Alice and Bob
can then estimate the maximal resolutions $\delta A_m^2$ and 
$\delta B_m^2$ that could have been
obtained by the eavesdropper. If these resolutions are 
sufficient to decode the information encoded in eigenstates
of $\hat{A}$ and $\hat{B}$, the line is not safe. 
On the other hand, security can be established if the 
noise levels given by the disturbance are low enough to
prevent the required measurement resolution. 

Another application of measurement uncertainties is the quantum
cloning problem. If it is known that the state to be cloned
is either an eigenstate of $\hat{A}$ or an eigenstate of 
$\hat{B}$, it is possible to define a quantitative cloning
error equal to the average quadratic deviation of the
clone's property $\hat{A}$ or $\hat{B}$ from the eigenvalue
of the original. This cloning error can then be used to
evaluate cloning strategies based on a quantum measurement
$\{\hat{M}_m\}$ on the original and a quantum state preparation
$\mid \psi_m\rangle$ for the clones. 
In this 
case, 
the disturbance caused by the measurement $\{\hat{M}_m\}$ 
characterizes the unavoidable damage done to the original in 
the cloning process, resulting in an irreversible loss of 
information about the original properties of the cloned state. 
If the cloning process extracts the maximal amount 
of information from the original system by effectively projecting
the system onto a pure state, it is also possible to define a 
set of cloning operators 
$\hat{C}=\mid \psi_m\rangle \langle \psi_m\mid$ for the 
optimal cloning procedure. 
This set of operators represents 
a projective measurement of the input system followed by a 
preparation of $N$ copies of the corresponding quantum state.
Note that it is possible to produce any number of clones
in this manner, since $\mid \psi_m\rangle$ is precisely defined 
by the classical measurement information $m$. The total output 
statistics of the cloning process is then given by a mixture of
the product states of $\mid \psi_m\rangle$ with the respective
statistical weight given by the  measurement probabilities
$p(m)$ for the original input state. However, the cloning 
errors for each individual clone can be estimated directly
from the disturbances caused by the cloning operator $\hat{C}$,
since it represents both the sensitivity of the cloning process
to the input and the resulting output statistics of all the
clones. 

Finally, it is also possible to apply this quantitative
characterization to errors and information extraction in
quantum teleportation. In this case, the measurement
made on the joint system of the input state and one part
of the entangled pair may be sensitive to properties of 
the unknown input state, e.g. because the entanglement
is non-maximal. This effect can be described by a set of
transfer operators $\hat{T}_m$ with properties equivalent 
to the measurement operators $\hat{M}_m$ 
\cite{Hof00,Hof01,Hof02}. 
The measurement resolution $\delta A_m^2$ then characterizes 
the information extracted about the input eigenvalue of 
$\hat{A}$, while the disturbance $\Delta B_m^2$ 
quantifies the teleportation error in $\hat{B}$.
A particularly simple example is given by the classical
limit of continuous variable teleportation, where a pair
of uncorrelated vacuum fields is used instead of the entangled
pair \cite{Bra98}. The teleportation procedure then corresponds to 
a measurement projection on a coherent state 
$\mid \alpha\rangle$, followed by the preparation of a
corresponding state in the output. This method can also be
used for quantum cloning or as an eavesdropping strategy.
In all cases, the procedure can be represented by the 
$\alpha$-dependent measurement operators
\begin{equation}
\hat{M}(\alpha) = \frac{1}{\sqrt{\pi}}
\mid \alpha \rangle\langle \alpha \mid.
\end{equation}
These operators can now be characterized using the 
quadrature components of the light field, $\hat{x}$ 
and $\hat{y}$, with $[\hat{x},\hat{y}]=i/2$, and 
the definitions of optimized estimates and uncertainties 
given by equations (\ref{eq:estimate}) and (\ref{eq:distall}). 
The results for the operators $\hat{M}(\alpha)$ then read
\begin{eqnarray}
x_\alpha + i y_\alpha &=& \alpha
\nonumber \\
\delta x^2_\alpha = \delta y^2_\alpha &=& 1/4
\nonumber \\
\Delta x^2_\alpha = \Delta y^2_\alpha &=& 1/2.
\end{eqnarray}
These uncertainties now define the noise levels in
the measurements and in the transmitted signal.
Note that the disturbances are twice as high as the 
measurement resolutions. This is a typical feature
of the classical teleportation limit \cite{Bra98}.
In an eavesdropping scenario, this strategy therefore
extracts maximal information but makes it easy 
for Alice and Bob to detect the eavesdropping 
attempt.

\section{Conclusions}

The effects of quantum measurements described by sets
of measurement operators $\{\hat{M}_m\}$ can be characterized in
terms of the physical properties of the measured system
by evaluating the effects of the measurement on eigenstates
of the corresponding Hermitian operators. 
It is then possible to define quantitative 
expressions for the concepts of measurement resolution 
and disturbance corresponding to the notions expressed
in the earliest discussions of quantum measurement 
\cite{Hei58}. 
These definitions allow a derivation of Heisenberg's 
uncertainty principle, demonstrating the general validity 
of uncertainty for all possible measurement strategies.
In particular, it can be shown that the back action of 
a generalized measurement is indeed uncertainty limited.
A complete characterization of generalized quantum 
measurements in terms of measurement resolutions 
and disturbances for each relevant physical property
may therefore provide practical insights into the 
nature of quantum measurements.

Since the definitions of measurement uncertainties have
been based on quantitative estimates of an unknown eigenstate
input, they can also be applied to evaluate errors in 
various quantum communication scenarios. For example,
eavesdropping strategies for quantum cryptography may require
an optimization of both the measurement resolutions
$\delta A_m^2$ and $\delta B_m^2$ for simultaneous estimates
of $\hat{A}$ and $\hat{B}$, and the corresponding
disturbances $\Delta A_m^2$ and $\Delta B_m^2$.
Similar considerations may also be useful in 
the discussion of quantum cloning and quantum teleportation.

\section*{Acknowledgements}
I would like to thank A. G. White for encouraging me to
write this paper and M. Ozawa and A. Furusawa for helpful 
comments on the problem of disturbance in quantum 
measurements.



\begin{appendix}
\section{Derivation of uncertainty relations for statistical
mixtures}

Although the basic derivation of uncertainty relations
for quantum states and density matrices is well known 
\cite{Nie}, it may be useful to review it
in the general context of statistical mixtures
in order to provide a more precise justification of
the measurement uncertainties discussed in this paper.

The basic derivation of uncertainty relations for
pure states is obtained from the Cauchy-Schwarz 
inequalities for the two Hilbert space vectors
given by
\begin{eqnarray}
&& (A_m-\hat{A})\mid \! \psi\rangle
\nonumber \\
\mbox{and} && (B_m-\hat{B})\mid \! \psi\rangle.
\end{eqnarray}
Since the product of the squared length of these
vectors must be larger or equal to the squared
inner product of the vectors, it follows
that
\begin{equation}
\langle \psi \mid (A_m-\hat{A})^2 \mid \psi \rangle
\langle \psi \mid (B_m-\hat{B})^2 \mid \psi \rangle
\geq |\langle \psi \mid (A_m-\hat{A})
     (B_m-\hat{B})\mid \psi \rangle|^2.
\end{equation}
The uncertainty relations are then obtained by taking
only the imaginary part of the inner product into 
account. Since $\hat{A}$ and $\hat{B}$ are Hermitian
operators, this imaginary part is given by one half
of the commutation relation, and the result is the 
well known formulation of uncertainty for pure states,
\begin{equation}
\underbrace{\langle \psi \mid (A_m-\hat{A})^2 \mid \psi 
\rangle}_{\delta A^2} \;
\underbrace{\langle \psi \mid (B_m-\hat{B})^2 \mid \psi 
\rangle}_{\delta B^2}
\geq \frac{1}{4}|\langle \psi \mid \; [\hat{A};\hat{B}]\; 
\mid \psi \rangle|^2.
\end{equation}
In order to generalize this result to density matrices 
or to any other form of statistical mixtures, it is 
sufficient to examine the case of a set of uncertainties
given by
\begin{equation}
\delta A_i^2 \delta B_i^2 \geq U_i^2,
\end{equation}
where the averaged uncertainties are given by 
\begin{eqnarray}
\delta A^2 &=& \sum_i p(i) \delta A_i^2
\nonumber \\
\delta B^2 &=& \sum_i p(i) \delta B_i^2.
\end{eqnarray}
It then follows that
\begin{eqnarray}
\delta A^2 \delta B^2 &=& \sum_{i,j} p(i)p(j) 
\frac{1}{2}\left(\delta A_i^2 \delta B_j^2 + 
\delta A_j^2 \delta B_i^2\right)
\nonumber \\
&\geq& \sum_{i,j} p(i)p(j) \delta A_i \delta B_i
                          \delta A_j \delta B_j
\nonumber \\
&\geq& (\sum_{i} p(i) U_i)^2.
\end{eqnarray}
It is therefore possible to derive an uncertainty 
relation for the statistical mixture defined by $p(i)$
by averaging over the uncertainties $U_i$.
This derivation can be applied to derive 
uncertainties for statistical operators
such as (\ref{eq:uncertainty1}) by representing the 
statistical operator as a mixture of pure states.
However, it can also be useful in a more general
context, as seen in the derivation of the back action
uncertainty (\ref{eq:uncertainty2}).

\end{appendix}


\begin{thebibliography}{xyz00}

\bibitem{Meas}
For an overview of fundamental problems in quantum measurement, see
{\it Quantum Theory and Measurement}, ed. by J.A. Wheeler and W.H. Zurek (Princeton University Press, Princeton 1983).
For the ongoing controversy regarding interpretation, see in 
particular the letters in Physics Today, {\bf 52}(2), 11 (1999), 
concerning the two part article by S. Goldstein, Physics Today 
{\bf 51}(3), 42 and {\bf 51}(4),38 (1998) and the letters in 
Physics Today, {\bf 52}(1), 15 (1999), concerning the article 
by M. Beller, Physics Today, {\bf 51}(9), 29 (1998).

\bibitem{EPR}
A. Einstein, B. Podolsky, and N. Rosen, Phys. Rev. {\bf 47}, 777 (1935).

\bibitem{Bel64} J.S. Bell, Physics (Long Island City, N.Y.) {\bf 1},195 (1964).

\bibitem{Mer90}
N.D. Mermin, Phys. Today {\bf 43} (6), 9 (1990).

\bibitem{Hei58} 
W. Heisenberg {\it Physikalische Prinzipien der Quantentheorie}
(S. Hirzel Verlag, Stuttgart 1958).

\bibitem{Nie}
M.A. Nielsen and I.L. Chuang, {\it Quantum Computation and
Quantum Information} (Cambridge University Press, Cambridge 2000),
p. 89.

\bibitem{Scu91}
M.O. Scully, B.G. Englert, and H. Walther, Nature {\bf 351},
111 (1991).

\bibitem{Art88}
E. Arthurs and M.S. Goodman, Phys. Rev. Lett. {\bf 60},
2447 (1988).

\bibitem{Oza02a}
M.Ozawa, Phys. Lett. A {\bf 299}, 1 (2002).

\bibitem{Oza02b}
M. Ozawa, quant-ph/0207121 (2002).

\bibitem{NieMT}
M.A. Nielsen and I.L. Chuang, {\it Quantum Computation and
Quantum Information} (Cambridge University Press, Cambridge 2000),
p. 84-86.

\bibitem{Bar00}
S. M. Barnett, D. T. Pegg, J. Jeffers, O. Jedrkiewicz, 
and R. Loudon, Phys. Rev. A {\bf 62}, 022313 (2000).  

\bibitem{Bar01}
S. M. Barnett, D. T. Pegg, J. Jeffers, and 
O. Jedrkiewicz, Phys. Rev. Lett. {\bf 86}, 2455 (2001). 

\bibitem{Dur98}
S. D\"urr, T.Nonn, and G. Rempe, Nature {\bf 395}, 33 (1998).

\bibitem{Ral00}
T.C. Ralph, Phys. Rev. A {\bf 62},
062306 (2000).

\bibitem{Hof01}
H.F. Hofmann, T. Ide, T. Kobayashi, and A. Furusawa,
Phys. Rev. A {\bf 64}, 040301 (2001).

\bibitem{Hof00}
H. F. Hofmann, T. Ide, T. Kobayashi, and A. Furusawa,
Phys. Rev. A {\bf 62}, 062304 (2000).

\bibitem{Hof02}
H.F. Hofmann, Phys. Rev. A {\bf 66}, 032317 (2002).

\bibitem{Bra98}
S.L. Braunstein and H.J. Kimble, Phys. Rev. Lett. {\bf80},
869 (1998).

\end{thebibliography}
\end{document}